\documentclass[prd,floatfix,nofootinbib,preprintnumbers,superscriptaddress,twocolumn]{revtex4}

\usepackage{epsf,epsfig,subfigure,graphicx,amsmath,amssymb}
\usepackage{amsfonts}
\usepackage{latexsym}
\usepackage{color}
\usepackage{accents}
\usepackage{hyperref}
\usepackage{verbatim}

\def\bea{\begin{eqnarray}}
\def\eea{\end{eqnarray}}

\def\Mp{M_{\rm pl}}
\def\Mpl{M_{\rm pl}}

\def\TeV{\,{\rm TeV}}
\def\GeV{\,{\rm GeV}}
\def\MeV{\,{\rm MeV}}
\def\keV{\,{\rm keV}}
\def\meV{\,{\rm meV}}

\begin{document}
\title{Dynamics of the cosmological relaxation after reheating}
\author{Kiwoon Choi}
\affiliation{Center for Theoretical Physics of the Universe, Institute for Basic Science (IBS), Daejeon, 34051, Korea}

\author{Hyungjin Kim}
\affiliation{Department of Physics, Korea Advanced Institute of Science and Technology, Daejeon, 34141, Korea}
\affiliation{Center for Theoretical Physics of the Universe, Institute for Basic Science (IBS), Daejeon, 34051, Korea}

\author{Toyokazu Sekiguchi}
\affiliation{Center for Theoretical Physics of the Universe, Institute for Basic Science (IBS), Daejeon, 34051, Korea}
\begin{abstract}
We examine if the cosmological relaxation mechanism, which was proposed recently as a new solution to the hierarchy problem, can be compatible with  high reheating temperature well above the weak scale.   As the barrier potential disappears at high temperature, the relaxion  rolls down further after  the reheating, which may ruin the successful implementation of the relaxation mechanism. It is noted that if the relaxion is coupled to a dark gauge boson, the new frictional force arising from dark gauge boson production can efficiently slow down the relaxion motion, which allows the relaxion to be stabilized after the electroweak phase transition for a wide range of model parameters, while satisfying the known observational constraints.
\end{abstract}

\preprint{CTPU-16-33}
\maketitle

\section{Introduction}
Cosmological relaxation of the Higgs mass has been proposed recently as an alternative solution to the weak scale hierarchy problem \cite{Graham:2015cka}.\footnote{See Refs. \cite{Antipin:2015jia,Gupta:2015uea,Espinosa:2015eda,Patil:2015oxa,Jaeckel:2015txa,Batell:2015fma,Matsedonskyi:2015xta,Marzola:2015dia,DiChiara:2015euo,Ibanez:2015fcv,Fonseca:2016eoo,Fowlie:2016jlx,  Evans:2016htp,Huang:2016dhp,Hardy:2015laa,Higaki:2016cqb,Kobayashi:2016bue,Choi:2016luu,Flacke:2016szy} for subsequent studies on the viability of the cosmological relaxation scenario.} In this scenario, a pseudo Nambu-Goldston boson $\phi$ is coupled to the Standard Model Higgs doublet $h$, scanning the Higgs mass from a large initial value to the small present value. This scalar field, often referred to as the relaxion, has  a potential including the piece enforcing the relaxion to move to scan the Higgs mass and also a periodic barrier potential to stop the relaxion at the position giving $m_h^2=-(89\, {\rm GeV})^2$. More specifically, the relevant scalar potential is given by 
\bea
\label{potential}
\big(\Lambda^2 - \frac{\Lambda^2}{f_{\rm eff}}\phi \big) |h|^2
-c_0\frac{\Lambda^4}{f_{\rm eff}} \phi +V_b,
\eea
where  $\Lambda$ denotes the Higgs mass cutoff scale, $f_{\rm eff}$ corresponds to the relaxion excursion required to scan the Higgs mass-square from $\Lambda^2$ to  $-(89\, {\rm GeV})^2$, $c_0$ is a positive coefficient of order unity as suggested by the naturalness argument,
and the barrier potential $V_b$ generically takes the form
$$
V_b=-\Lambda_b^4(h)\cos(\phi/f)
$$
with a Higgs-dependent amplitude 
$$
\Lambda_b^4(h)=\mu_b^{4-n}h^n,
$$  where
$\mu_b$ is determined by the scale where $V_b$  is generated, as well as by the involved couplings.
Imposing the stationary condition to the potential (\ref{potential}), one finds  
\bea
\label{stationary}
\frac{f_{\rm eff}}{f} \,\sim\,  \frac{\Lambda^4}{\Lambda_b^4(h=v)}\frac{1}{\sin(\phi_0/f)},
\eea
where $v=246$ GeV and $\sin(\phi_0/f)\sim v^2/(v^2+\Lambda_b^2)$ \cite{Choi:2016luu} for the relaxion vacuum value $\phi_0$ in the present universe.

There can be two ways to generate the barrier potential as discussed in the literatures \cite{Graham:2015cka,Antipin:2015jia,Gupta:2015uea}.  The minimal scenario is to generate $V_b$  through the relaxion coupling to the QCD anomaly, i.e.
$\phi G\tilde G/32\pi^2 f$,  which would result in $\Lambda_b^4 \sim y_u h \Lambda_{\rm QCD}^3$, where $\Lambda_{\rm QCD}\sim 200$ MeV is the QCD scale and $y_u\sim 10^{-5}$ is the up-quark Yukawa coupling. In this case, $\phi_0/f$ is identified as the QCD vacuum angle $\theta_{\rm QCD}$ and therefore  constrained as $|\sin(\phi_0/f)|\lesssim 10^{-9}.$
Alternatively, the barrier potential can be generated by a new physics around the weak scale, yielding
for instance $\Lambda_b^4 = \mu_b^2 |h|^2$ with $\mu_b$ around the weak scale.

To implement the relaxation mechanism, the amplitude  of the barrier potential is required to be bounded as
$\Lambda_b^4(h=v)\lesssim {\cal O}(16\pi^2 v^4)$ \cite{Graham:2015cka,Antipin:2015jia,Gupta:2015uea,Espinosa:2015eda}, where $v=246$ GeV is the Higgs vacuum value in the present universe. Then the stationary condition (\ref{stationary}) shows that the relaxion mechanism transmutes the weak scale hierarchy
$\Lambda \gg v$ to another hierarchy $f_{\rm eff}\gg f$. Although the latter hierarchy can be technically natural, it may require an explanation for its origin. This issue has been addressed in \cite{Choi:2015fiu,Kaplan:2015fuy}, proposing
a scheme to generate an exponential hierarchy $f_{\rm eff}/f\sim e^N$ based on models with $N$ axions \cite{Kim:2004rp,Choi:2014rja}.  

A key ingredient of the relaxation scheme is a mechanism to dissipate away the relaxion kinetic energy which is originating from 
the initial potential energy of ${\cal O}(\Lambda^4)$. It is usually assumed that the relaxion loses its kinetic energy 
by  the Hubble friction during the inflationary period. Then the scheme requires a rather large number of inflationary $e$-foldings \cite{Graham:2015cka}, which is estimated as
\cite{Choi:2016luu}
\bea
\label{efolding_1}
N_e  \sim  \frac{\Lambda^4}{\Lambda_b^4}\left(\frac{v^2+\Lambda_b^2}{v^2}\right)^2
\eea
for the case that the barrier potential is induced by new physics, and
\bea
\label{efolding_2}
N_e \sim \frac{\Lambda^4}{\theta_{\rm QCD}y_uv\Lambda_{\rm QCD}^3} \gtrsim 10^{24}\left(\frac{\Lambda}{{\rm TeV}}\right)^4
\eea
for the other case that the barrier potential is induced by low energy QCD.
The above result and the relaxion scale hierarchy (\ref{stationary}) show  that the scenario with QCD-induced
barrier potential requires a huge $e$-folding number and also a  big hierarchy among the relaxion scales. As a too large $e$-folding number might cause a severe fine-tuning problem, in the following we will  focus on the scenario that
the barrier potential is generated by new physics around the weak scale, yielding
\bea
\label{barrier}
V_b=-\Lambda_b^4(h)\cos(\phi/f)=-\mu_b^2 |h|^2\cos(\phi/f)
\eea
with $\mu_b\lesssim {\cal O}(4\pi v)$. By the same reason, we will be more interested  in the case that $\mu_b$ is somewhat close to the weak scale.

In the relaxation scenario, to avoid a fine-tuning of the initial condition, the relaxion is assumed to be stabilized before the inflation is over. 
If the temperature during the reheating phase is well below the weak scale 
which corresponds to the scale where the barrier potential is generated, the relaxion dynamics after the reheating  is trivial. It stays there without changing the Higgs mass selected during inflation.
However, if the universe experiences a high temperature $T\gg v$ after  inflation,  the electroweak gauge symmetry is restored and the barrier potential disappears. Then the relaxion starts to roll again until the temperature cools down  to a critical temperature $T_c\sim v$ where the barrier potential is developed again, and such subsequent evolution may ruin the successful implementation of the relaxation mechanism.  
On the other hand, high reheating temperature $T_R \gg v$ is often favored for viable cosmology, in particular for baryogenesis. It is therefore an interesting question if the cosmological relaxation mechanism can be compatible with such high reheating temperature.
In this paper, we wish to examine such possibility within the relaxion scenario in which the barrier potential (\ref{barrier}) is generated by new physics near the weak scale\footnote{
The possibility of high reheating temperature was discussed in 
\cite{Graham:2015cka} for the case of QCD-induced barrier potential.}.

To proceed, let us first consider the case that there is no additional dissipation of  the relaxion energy other than those by the Hubble friction. During the period when $V_b$ is negligible, e.g. for the radiation-dominated period with $T > v$, solving the equation of motion determined by (\ref{potential}),  one finds that 
the relaxion speed behaves as
\bea
\dot{\phi}(t) \,\simeq\, \frac{\Lambda^4}{f_{\rm eff}}t \,\simeq\, \frac{\Lambda_b^4}{f}\left(\frac{90}{4\pi^2 g_*(T)}\right)^{1/2}\frac{M_{\rm pl}}{T^2},
\eea
where $\Lambda^4_b \equiv \Lambda^4_b(h=v)=\mu_b^2 v^2$, $g_*(T)$ is the number of relativistic degrees of freedom at $T$, and we 
 use the relation (\ref{stationary}) for the last expression.
As the relaxion speed is increasing in time, to stop the relaxion by the barrier potential 
developed around the time $t_c$,
one needs 
\bea
\label{condition_1}
\dot{\phi}(t_c) \lesssim \Lambda_{b}^2.
\eea
One can also make sure that if this condition is satisfied, the relaxion is successfully stabilized within a few Hubble time
from $t_c$ with a total excursion $\Delta \phi \lesssim {\cal O}(f)$, and therefore the corresponding change of the Higgs mass is
negligible.
On the other hand, the condition (\ref{condition_1}) puts a lower bound on the relaxion decay constant $f$, given by 
\bea
\label{bound_f}
\frac{f}{\Mp} \,\gtrsim\, \left(\frac{90}{4\pi^2 g_*(v)}\right)^{1/2}\frac{\Lambda_{b}^2}{v^2},
\eea
where we used $T_c\sim v$.
Although it is possible that the dynamics to generate $V_b$ involves  a small coupling, so  $\Lambda_b \ll v$, such a small $\Lambda_b$ is disfavored as it requires a bigger $e$-folding number (\ref{efolding_1}) for a given value of the Higgs mass cutoff $\Lambda$.
For $\Lambda_b\sim v$ which is more favored in view of (\ref{efolding_1}), the  bound (\ref{bound_f}) suggests that $f$ should be at least
comparable to the Planck scale.

The above observation implies that one needs an additional mechanism to dissipate  the relaxion energy to make the scheme compatible with
 $T_R\gg v$ for the more interesting  parameter range with 
$\Lambda_b\sim v$ and $f\ll M_{\rm pl}$.  
It is well known that a rolling scalar field $\phi$ can lose its kinetic energy through gauge field production
induced by the coupling,
\bea
\frac{1}{4}\frac{\phi}{{F}} X_{\mu\nu} \widetilde{X}^{\mu\nu} ,
\label{anomalous_coupling}
\eea
where $X_{\mu\nu}=\partial_{[\mu}X_{\nu]}$  is an Abelian gauge field strength  and 
$\widetilde X_{\mu\nu}$ is its dual.
In the presence of this coupling, a rolling  $\phi$ develops tachyonic instability of $X_\mu$, causing an exponential  growth of $X_\mu$ for certain range of wave number.
 This provides an effective frictional force to the motion of $\phi$, which has been applied recently to 
 the relaxion dynamics in the early universe \cite{Hook:2016mqo}.\footnote{Identifying $X_\mu$ as the electroweak gauge bosons
 ($W^a_\mu$ for $SU(2)_L$ and $B_\mu$ for $U(1)_Y$) whose masses are determined by the Higgs vacuum value, Ref. \cite{Hook:2016mqo} argued that a particular form of the coupling \eqref{anomalous_coupling}, i.e. $\phi\left(g^2 W^a_{\mu\nu}\tilde W^{a\mu\nu}-g^{\prime 2} B_{\mu\nu}\tilde B^{\mu\nu}\right)$, can provide Higgs-depedent back reaction to relaxion evolution, stabilizing the relaxion field at the desired value giving $\langle h \rangle = v$.}
In this paper, we explore the possibility of high reheating temperature in the relaxation scenario in which the coupling
(\ref{anomalous_coupling}) is mainly responsible for the relaxion energy dissipation after reheating. 
We focus only on the dynamics of relaxion after reheating, by assuming that the electroweak scale is already selected by relaxion during the inflationary period as in the original cosmological relaxion scenario \cite{Graham:2015cka}. 
As we will see,
in case that $X_\mu$ is identified as the $U(1)_Y$ gauge boson of the Standard Model (SM), due to its large thermal mass, the gauge field production  is not efficient enough to slow down the relaxion motion in most of the parameter space allowed by other constraints.  
On the other hand, if $X_\mu$ is identified as a dark gauge boson with negligible thermal mass, the gauge field production 
can efficiently slow down the relaxion motion, allowing the relaxion to be successfully  stabilized after the electroweak phase transition for a wide range of model parameters.

This paper is organized as follows. In Sec.~\ref{sec:production_gauge_field}, we review the gauge field production by a rolling scalar field with the coupling (\ref{anomalous_coupling}), and apply the results for the relaxion dynamics at $T\gg v$. Our primary concern is to identify the parameter region which allows the reheating temperature $T_R\gg v$ without modifying the standard cosmology after reheating.
For this, we estimate the relaxion excursion and terminal speed  at the time when the relaxion is stopped by a barrier potential developed at $T_c\sim v$. We provide also numerical results to cross check our analytic estimation. 
In Sec.~\ref{sec:discussion}, we discuss additional constraints on the scenario discussed in the previous section, and Sec.~\ref{sec:conclusion} is the conclusion.

\section{Relaxion dynamics with gauge field production}\label{sec:production_gauge_field}

In the presence of the coupling (\ref{anomalous_coupling}),
a background evolution of relaxion  develops tachyonic instability of the Abelian gauge boson $X_\mu$ \cite{Anber:2009ua,Notari:2016npn}. 
Let us begin with the case without any light $U(1)_X$ charged particle, in which the gauge field production turns out most efficient.
In this case, there is no thermal mass of $X_\mu$ even at high temperature limit,  and then the equation of motion for $X_\mu$  in the expanding universe is given by
\bea
X''_\pm + \left( k^2 \mp a k \frac{\dot \phi}{{F}} \right) X_\pm = 0,
\eea
where $\pm$ denotes the helicity state, $a$ is the scale factor of the expanding universe with the metric 
$$ds^2= dt^2-a^2(t)dx^2=a^2(\tau)\left(d\tau^2-dx^2\right),$$ and
$X^\prime = dX/d\tau$ and $\dot\phi=d\phi/dt$ for the conformal time $\tau$ and the physical time $t$. {Assuming $\dot{\phi}>0$, the positive helicity state experiences tachyonic instability for the wave number $k \leq k_{\rm max} = (a\dot{\phi}/{F})$.} Using the WKB approximation, we find that the corresponding gauge field modes grow as 
\bea
X_+(k) \sim
\frac{1}{\sqrt{2k}} \exp\left[ \int^\tau  d\tau' \, \Omega(k,\tau')\right],
\eea
with the growth rate determined as
\bea
\Omega^2= ak \frac{\dot{\phi}}{F} - k^2
\label{tachyon_mass1}
\eea
for $F(\ddot{\phi} + H \dot{\phi})^2 / \dot{\phi}^3\leq k/a \leq \dot{\phi}/{F}$. Here the lower bound on $k$ is required for the validity of WKB approximation, $|\Omega'/\Omega^2| \ll 1$. Note that the gauge field production is dominated for 
the modes with $k\sim k_{\rm max}$.

\begin{figure}
\includegraphics[scale=0.24]{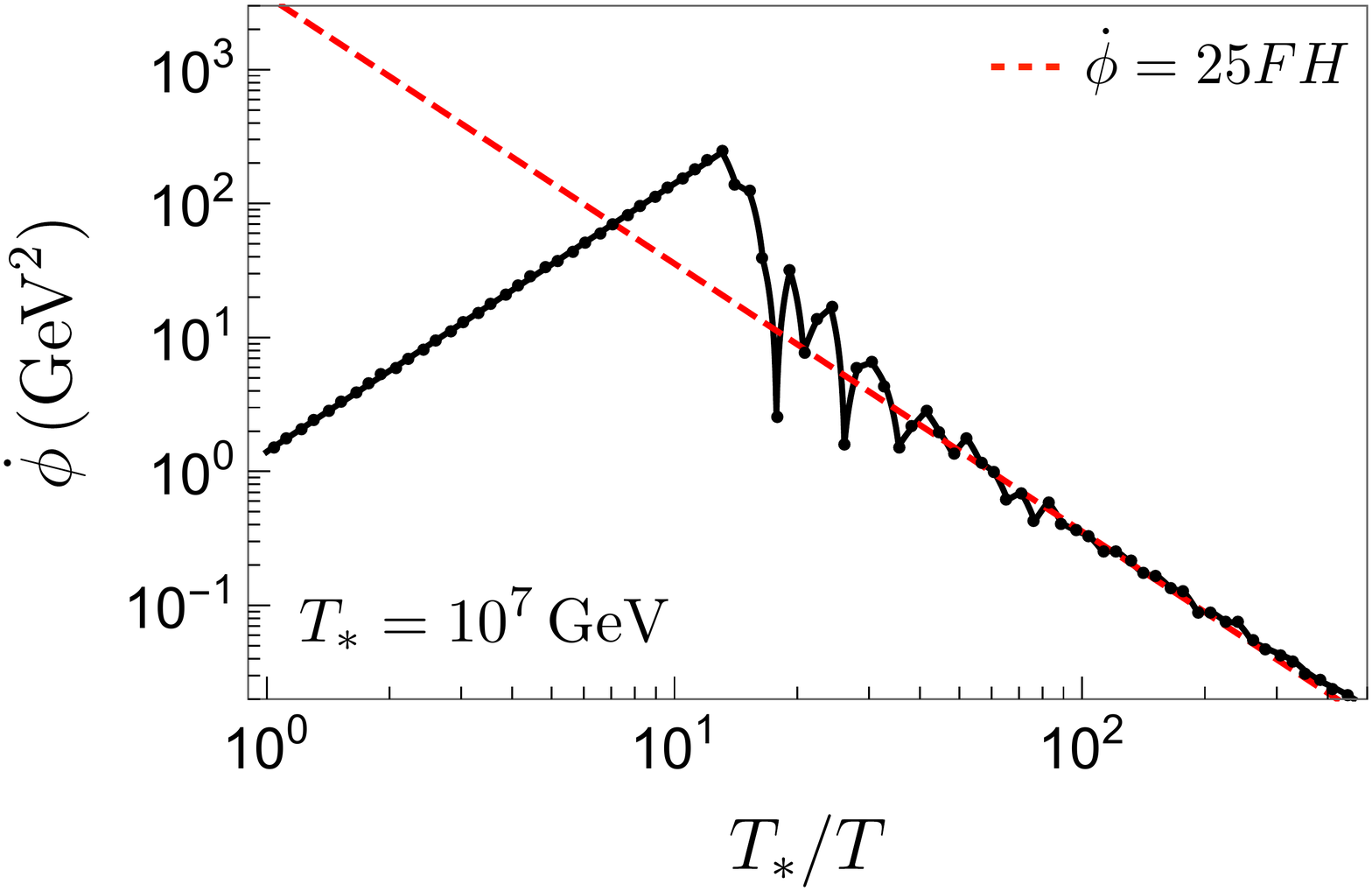}
\caption{The evolution of relaxion speed for the model parameters of Table~\ref{tab:parameter_set_1}, when there is no $U(1)_X$ charged particle in the background thermal plasma. The relaxion speed increases as $\dot\phi \simeq \Lambda_b^4/5fH$, and approaches $\dot{\phi}_{\rm term} \simeq \xi FH$ with $\xi \simeq 25$ when the temperature of the universe is lower than $10^6 \GeV$. The initial temperature is taken to be  $T_* = 10^7 \GeV$. }
\label{fig:vel_no_plasma}
\end{figure}

\begin{figure}
\centering
\includegraphics[scale=0.24]{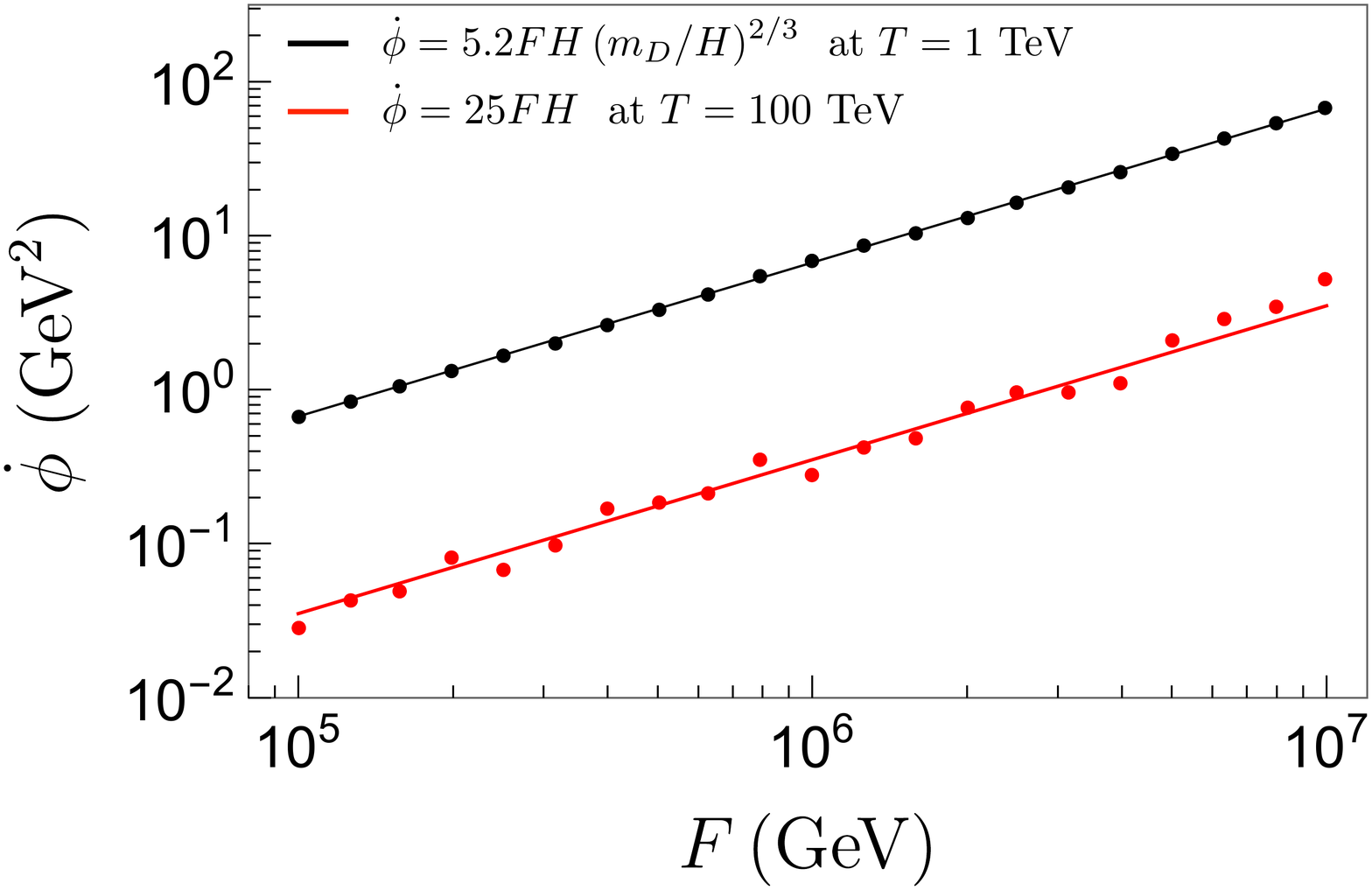}
\caption{The black line corresponds to the relaxion terminal speed at $T= 1 \TeV$ when there is 
dark plasma with temperature $T_d =10^{-5}g^\prime T/g_X$, providing a thermal mass of $X_\mu$ bigger than the Hubble expansion rate, while the red line is  the terminal speed at $T = 100 \TeV$ in the absence of dark plasma.  The model parameters chosen here are described in Table~\ref{tab:parameter_set_1}. These numerical results are well matched to the expression  $\dot{\phi} =\tilde{\xi} FH(m_D/ H)^{2/3}$ with $\tilde{\xi} \simeq 5.2$, and ${\dot \phi}= \xi F H$ with $\xi \simeq 25$. }
\label{fig:terminal_velocity}
\end{figure}

So a rolling relaxion produces gauge fields with comoving wave number $k\leq k_{\rm max}$, and  those gauge fields will eventually modify the evolution of relaxion.
To see the interplay between the gauge field production and the relaxion evolution, we consider the relaxion equation of motion 
\bea
\frac{d \dot{\phi}}{dt} = - 3 H \dot{\phi} - \frac{\partial V}{\partial \phi} - \frac{1}{4 {F} a^4} \big\langle X_{\mu\nu} \widetilde{X}^{\mu\nu} \big\rangle,
\label{eom_scalar}
\eea
where the ensemble average of gauge fields is given by
\bea
\frac{1}{4{F}} \big\langle X_{\mu\nu} \widetilde{X}^{\mu\nu} \big\rangle 
&=& 
\frac{1}{4F\pi^2} \int dk\, k^3 \frac{d}{d\tau} \left( | X_+|^2 - |X_-|^2 \right)
\nonumber\\
&\propto& \exp \bigg[ 2\int^\tau d\tau' \, \Omega(k_{\rm peak}, \tau') \bigg],
\label{gauge_friction}
\eea
where $k_{\rm peak}(\tau)=k_{\rm max}(\tau)/2$ is the wave length at which the gauge field production is maximized. The above equation of motion  shows that the produced gauge fields provide additional frictional force to the relaxion motion.
Then the relaxion speed reaches at its terminal value  around the time when the accelerating force $\partial V / \partial \phi$ in \eqref{eom_scalar} is balanced by 
the last frictional force term. We already noticed  that the gauge field production is most efficient for $k\sim k_{\rm max}$. Then, equating \eqref{gauge_friction} with $\partial V / \partial \phi$, the relaxion  terminal speed is estimated as
\bea
\dot{\phi}_{\rm term}= \xi H F,
\label{term_v}
\eea
where the dimensionless coefficient $\xi$ mildly depends on various factors such as $(\partial V / \partial \phi)$, ${F}$, and the initial condition for $X_\mu$. We performed a numerical analysis to examine the relaxion evolution for the model parameters in Table~\ref{tab:parameter_set_1}, and depict the result in 
Fig.~\ref{fig:vel_no_plasma}. Our result shows that the relaxion speed indeed approaches the form (\ref{term_v})
with $\xi \simeq  25$ as indicated in Fig.~\ref{fig:terminal_velocity}.

\begin{table}[h]
\centering
\begin{tabular}{|c|c|c|c|c|}
\hline
$\Lambda$ & $\Lambda_{b}$ & $f_{\rm eff}$ &$f$ & ${F}$
\\
\hline
$ 10^4 \GeV$ & $10 \GeV$ & $10^{19} \GeV$ & $10^7 \GeV$ & $10^6 \GeV$ 
\\
\hline
\end{tabular}
\caption{Sample model parameters for numerical analysis.}
\label{tab:parameter_set_1}
\end{table}

We can also estimate the time scale of gauge field production. 
Right after the reheating, the relaxion field begins to roll down with a speed
\bea\label{relaxion_no_gauge}
\dot\phi 
\simeq \frac{2}{5}\frac{\Lambda_b^4}{f}t [1-(t_R/t)^{5/2}],
\eea
where $t_R$ denotes the time of reheating.
Gauge field production  due to this relaxion motion
becomes important when 
$$ \int^\tau d\tau' \, \Omega(k_{\rm peak}, \tau') = {\cal O}(1),$$ 
for which the frictional force term $\propto \langle X\tilde X\rangle$
in (\ref{eom_scalar}) is  not negligible anymore.
Imposing this condition to the solution (\ref{relaxion_no_gauge}),  we find the gauge field production time scale is  given by\footnote{If the inflationary Hubble scale $H_I < 1/t_p$, the friction from gauge field production dominates over the Hubble friction during inflationary period,  which would require even larger   inflationary $e$-folding number for the scanning of the Higgs mass.} 
\bea
t_p \sim \sqrt{ f F / \Lambda_b^4 }.
\eea

Soon after $t_p$, the relaxion speed approaches its terminal value given by
(\ref{term_v}). Since this terminal speed is decreasing in time, the relaxion field keeps losing its kinetic energy, and its speed eventually becomes smaller than the height of the barrier potential developed at $T=T_c\sim v$. Assuming that the universe is radiation-dominated over the period under consideration, we estimate the temperature $T_b$ when $\dot{\phi}^2(T_b) = \Lambda_b^4$ as
\bea
T_b = \left(\frac{90}{\pi^2 g_*(T_b)} \right)^{1/4} \left( \frac{\Mp \Lambda_b^2}{\xi F} \right)^{1/2}.
\eea
If $T_b> T_c$,  $\dot\phi(T_c)$ is small enough to be stopped by the barrier potential right after the barrier potential is developed at $T_c\sim v$. On the other hand, if $T_b < T_c$, the relaxion rolls until the lower temperature $T_b$ when its speed is further reduced down to $\Lambda_b^2$.
Then the temperature when the relaxion is finally stabilized  is given by
\bea
T_s = \min(T_b, T_c) =\min (T_b, v),
\eea
where we set $T_c=v$ for simplicity.

Having determined the terminal speed, we can now compute the relaxion excursion
after the reheating, which is given by\footnote{Note that the reheating temperature $T_R$ usually means the temperature when the reheating  is completed, which is given by $T_R \simeq 1.7 g_*^{-1/4}\sqrt{\Mpl \Gamma_{\sigma}}$ where $\Gamma_{\sigma}$ is the inflaton decay width, and $g_*$ is the number of relativistic degrees of freedom at $T_R$. On the other hand, the relaxion experiences a subsequent rolling if the maximal temperature during the reheating period, which is given by
$T_{\rm max}\sim T_R(H_{\rm end}/H(T_R))^{1/4}\gg T_R$, is higher than the weak scale.
As the relaxion terminal speed and the excursion range are almost independent of the initial temperature, we ignore the difference between $T_R$ and $T_{\rm max}$.} 
\bea
\Delta \phi \simeq  
\frac{1}{2} \xi F \ln(H(t_p)/H(t_s)), \label{excur}
\eea
where $t_s$ is the time when the relaxion is finally stabilized, i.e. when $T=T_s$.

The dark gauge bosons produced by rolling relaxion eventually contribute to the dark radiation at the time of the big bang nucleosynthesis (BBN). Imposing the observational bound on dark radiation, $\Delta N_{\rm eff}\lesssim 0.3$, the energy density of $X$ gauge bosons at $t_s$ is bounded roughly as
\bea
\rho_X(t_s)
\,\simeq\, \frac{1}{a^4(t_s)} \int_{t_p}^{t_s} dt' a^4(t') \dot{V}
\,\simeq\, \Lambda_b^4 \frac{\xi F}{4f}
\,\lesssim\, T_s^4
.\eea 
Note that this bound from dark radiation ensures that the universe is radiation-dominated over the period from the beginning of reheating to the re-stabilization of relaxion.
From this condition, we finally find a lower bound on $f/F$ as
\bea
\frac{f}{F} \,\gtrsim\,  \frac{\xi}{4}\frac{\Lambda_b^4}{T_s^4},
\label{bound}
\eea
where $\xi\simeq 25$.

\begin{figure*}[t]
\centering
\begin{tabular}{cc}
\includegraphics[scale=0.255]{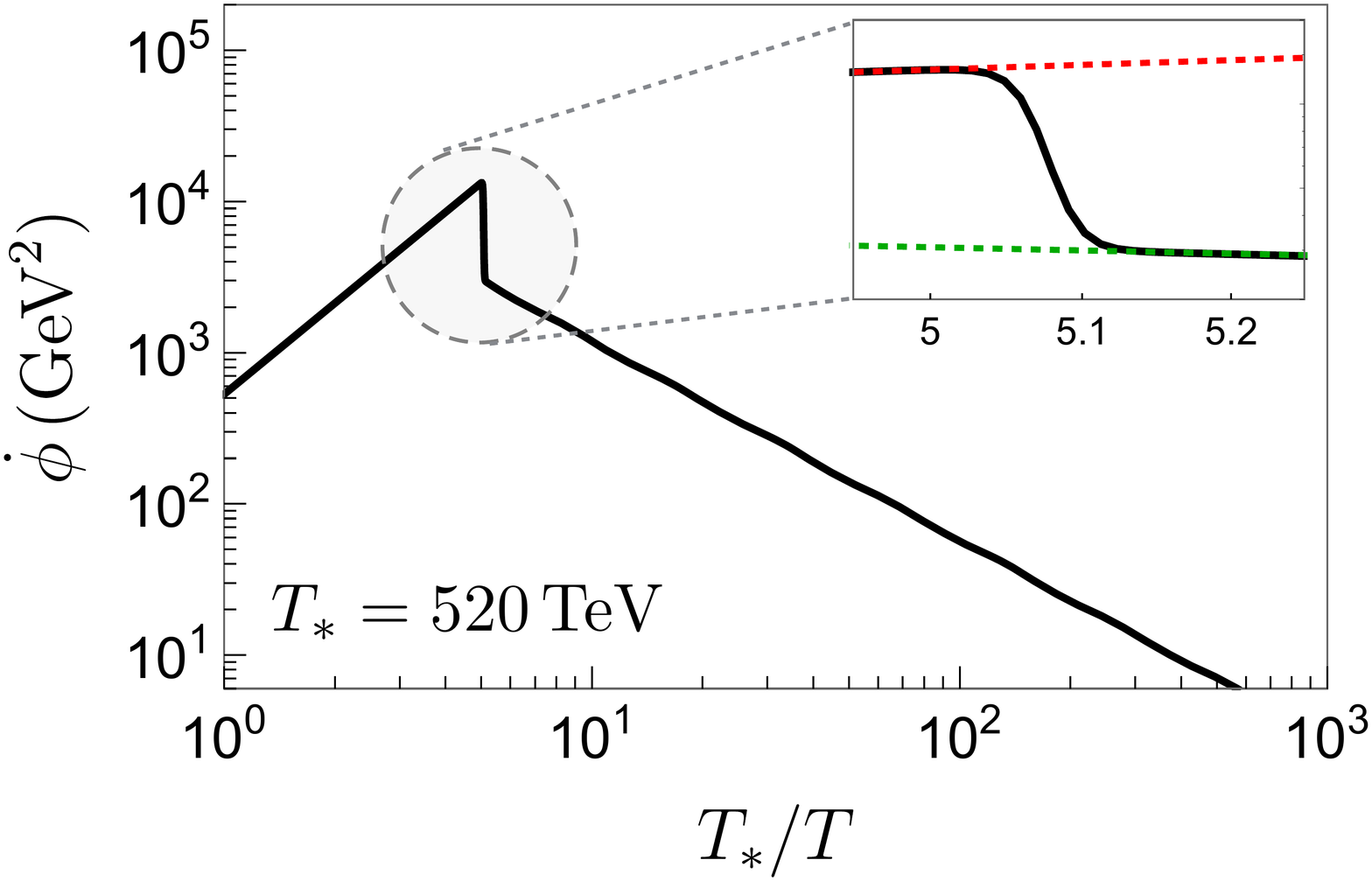}\hspace{1.5cm} &
\includegraphics[scale=0.44]{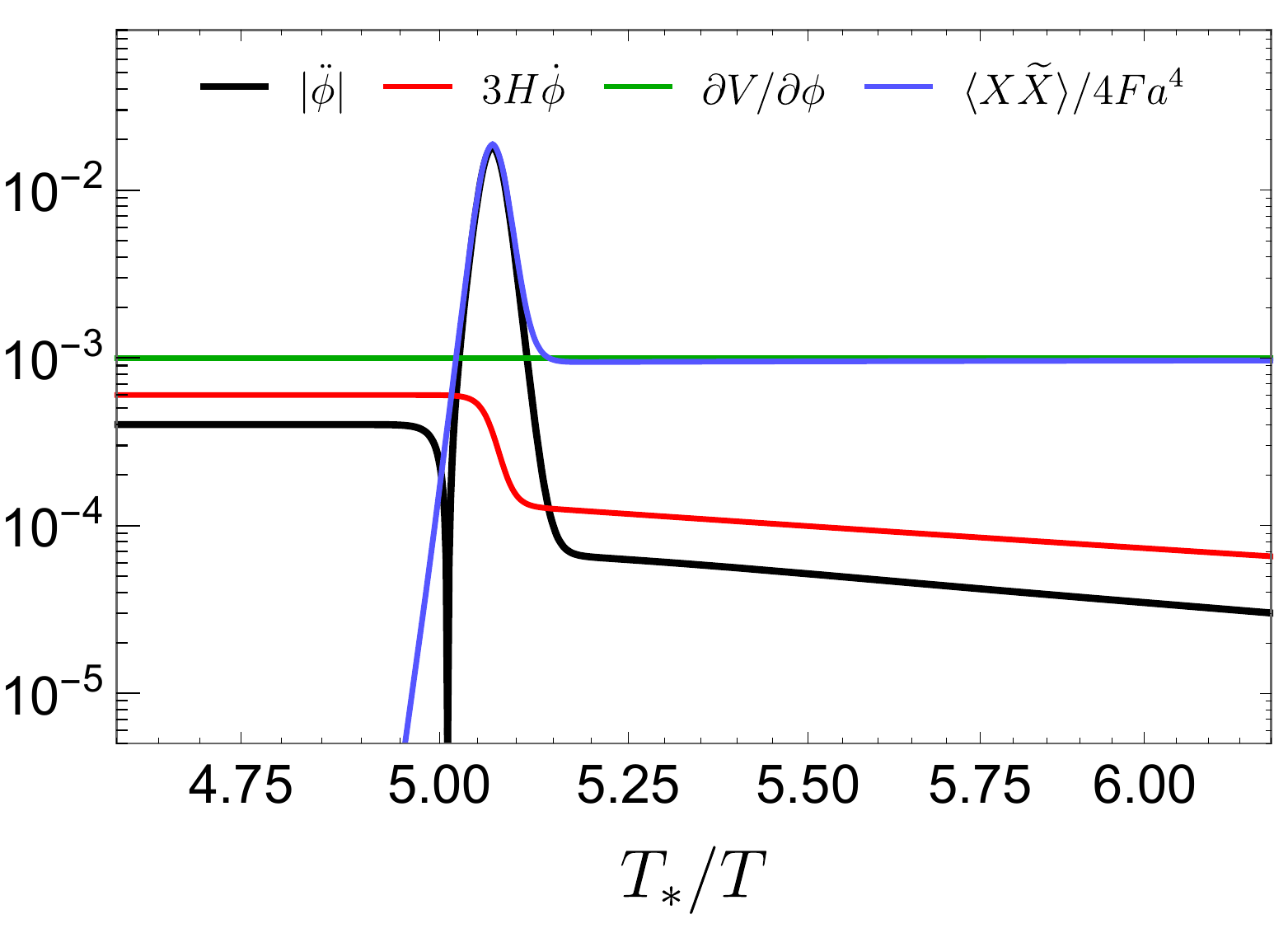}\hspace{1cm}
\end{tabular}
\caption{(Left) The evolution of relaxion speed when $X_\mu$ has a thermal mass bigger than the Hubble expansion rate. The initial temperature is set as  $T_* = 520 \TeV$.  Here the red dotted line represents  $\dot{\phi} = \Lambda_{b}^4 / 5 f H$, while the green dotted line represents $\dot{\phi}_{\rm term} = \tilde{\xi} FH (m_D/H)^{2/3}$
with $\tilde\xi=5.2$. (Right) Evolution of the four
quantities in the relaxion equation of motion (\ref{eom_scalar}).
}
\label{fig:vel_eom}
\end{figure*}

Up to this point, we assumed that there is no light $U(1)_X$ charged particle in the thermal plasma
to ensure that we can safely ignore the thermal mass of $X_\mu$.
Having a thermal mass higher than the Hubble expansion rate can significantly change the result as it suppresses the gauge field production, so makes the mechanism less efficient.
To examine this issue, let us assume that there is a SM singlet but $U(1)_X$ charge particle constituting a thermal plasma with temperature $T_d < T$, where
$T$ is the temperature of the SM degrees of freedom. 

In the presence of dark plasma, the dispersion relation of $X_\mu$ is changed as
\bea
\omega^2 - k^2 + ak \frac{\dot \phi}{F} = G(\omega,k),
\eea
with the one loop thermal correction given by \cite{Kapusta:2006pm,Bellac:2011kqa}
\bea
G(\omega,k) = m_D^2 \frac{\omega}{k} \left[ \frac{\omega}{k} + \frac{1}{2} \bigg\{ 1 - \bigg( \frac{\omega^2}{k^2} \bigg) \bigg\} \ln \frac{ \omega + k }{ \omega - k} \right].
\eea
Here $m_D$ is the Debye mass of $X_\mu$  given by 
$$m_D^2 = \frac{g_X^2 T_d^2}{6},$$ where $g_X$ is the $U(1)_X$ gauge coupling. Since Abelian gauge boson does not have a magnetic mass, the dispersion relation still allows tachyonic modes for $\Omega\ll k<k_{\rm max}$. However, contrary to the previous case, the tachyonic instability is alleviated by the thermal mass, which results in the tachyonic gauge field modes grow as
\bea
X_+(k) \sim \frac{1}{\sqrt{2k}} \exp\left[ \int^\tau d\tau' \, \Omega(k, \tau') \right]
\eea
with a reduced growth rate:
\bea
\frac{\Omega}{a} = \frac{(k/a)^2 }{m_D^2} \frac{\dot\phi}{F},
\label{tachyon_mass2}
\eea
where we assumed  
$$
H < \Omega/a <k/a< m_D.
$$ If $T_d$ is small enough to yield $m_D<H$, the gauge field growth rate is  approximately given by \eqref{tachyon_mass1}, and therefore essentially same as the case without any light $U(1)_X$ charged particle.

As in the previous case, we can estimate the gauge field production time scale and the relaxion terminal speed for the case with a thermal mass $m_D > H$.
We then find 
\bea
t_p \sim  \left( \frac{\kappa^2g'^2 }{\sqrt{g_*(T_p)}} \frac{\Mp f^3 F^3}{\Lambda_{b}^{12}}\right)^{1/5}.
\label{production_time}
\eea
and
\bea
\dot{\phi}_{\rm term}\simeq \tilde{\xi} F H (m_D/H)^{2/3},
\label{term_plasma}
\eea
where $\kappa \equiv g_XT_d/g' T$, $m_D=g_XT_d/\sqrt{6} = \kappa g' T/\sqrt{6}$ and the numerical coefficient  $\tilde\xi\simeq 5.2$ (see Fig.~\ref{fig:terminal_velocity}).
As we  mentioned in the previous paragraph, this result is valid only when $m_D> H$. If $m_D<H$,  the terminal speed should be replaced with \eqref{term_v}. Note that 
compared to  \eqref{term_v}, the terminal speed is bigger by the factor $(m_D/H)^{2/3}$, showing that a gauge field thermal mass $m_D \gg H$  makes the gauge field production much less efficient, yielding a much bigger value of the final relaxion speed.
 In Fig.~\ref{fig:vel_eom}, we depict the evolution of relaxion speed, as well as the evolution of the four quantities that appear in the relaxion equation of motion (\ref{eom_scalar}) for the model parameters in Table~\ref{tab:parameter_set_1}
 and the gauge field thermal mass $m_D=g_X T_d/\sqrt{6}=10^{-5}g'T/\sqrt{6}$ \footnote{Here we assume that $\kappa=g_X T_d/g'T$ is constant during the evolution of relaxion field. What the rolling relaxion field mostly produces is a magnetic component of dark gauge fields. Contrary to the electric component of dark gauge boson, the magnetic component is not thermalized if $k_{\rm max} < k_{\rm diff}$ where $k_{\rm diff}$ is a diffusion scale given by $k_{\rm diff}/a \sim \sqrt{T_d H /\alpha_X}$  $(\alpha_X=g_X^2/4\pi)$ \cite{Grasso:2000wj}. Throughout this paper, we consider $g_X \sim {\cal O}(10^{-1})$ so that $k_{\rm max}<k_{\rm diff}$ for a relevant range of parameter space. In this case, the produced gauge fields do not heat up the dark sector temperature.}.

Although the rate is slower than the case without thermal mass, the terminal speed  is again decreasing in time.
  The temperature of the SM particles when the terminal speed becomes comparable to $\Lambda_b^2$ is estimated as
\bea
T_b \sim 
\frac{1}{\sqrt{g'\kappa}}
\left( \frac{90}{\pi^2 g_*(T_b)} \right)^{1/8}
\left( \frac{\Mp \Lambda_b^6}{\tilde{\xi}^3 F^3} \right)^{1/4},
\eea
and then the relaxion is finally stabilized at the temperature $T_s = \min(T_b, v)$. 
Having determined  $T_s$, it is straightforward to find the relaxion excursion after the reheating, which is given by
\bea
\Delta \phi = \left.\tilde{ \xi} F(m_D/H)^{2/3}\right|_{T=T_s}.
\label{excur_plasma}
\eea
We can impose again the constraint 
$$\rho_X(T_s) \,\sim\, \frac{\Lambda_b^4}{f}\Delta \phi \,\lesssim\, T_s^4
$$
to avoid too much dark radiation, which yields
\bea
\frac{f}{F} \,\gtrsim\, 
{\tilde{\xi}} 
\frac{\Lambda_b^4}{T_s^4}
\left( \frac{m_D}{H}\right)^{2/3} \bigg|_{T=T_s}.
\label{bound_plasma}
\eea

Generically an Abelian dark gauge boson $X_\mu$ can have a kinetic mixing with
the $U(1)_Y$ gauge boson $B_\mu$ in the SM:
$$
\Delta {\cal L} \,=\, \epsilon X_{\mu\nu} B^{\mu\nu},
$$
which may result in a modification of our results, as well as rich phenomenological
consequences as discussed in \cite{Kaneta:2016wvf}.
In fact, after proper diagonalization of the kinetic and thermal mass terms, we find that the modification due to the kinetic mixing is suppressed by $\epsilon^2 (k\dot{\phi} / F)/T^2$, and therefore can be safely ignored.

Let us finally remark the possibility that $X_\mu$ is identified as 
the $U(1)_Y$ gauge boson in the SM.
The mechanism of gauge field production can hardly be realized under the working assumption of this paper if $X_\mu$ is the hypercharge gauge boson. The above discussions can be directly applied to the hypercharge gauge boson by setting $\kappa = 1$. As is already shown in \cite{Choi:2016luu,Flacke:2016szy}, the relaxion window for $m_\phi \gtrsim {\cal O}(0.1 \rm MeV)$ is strongly constrained by rare meson decay, electric dipole moment, and supernova 1987A. Only tiny window for the relaxion is available. Instead, we could focus on relatively light relaxion mass, $m_\phi \leq {\cal O}(0.1 \rm MeV)$, while requiring
$$
F \gtrsim 10^{10} \, {\rm GeV}
$$
to satisfy the bound on the relaxion-photon coupling \cite{Jaeckel:2010ni}. For this size of coupling strength, the relaxion will be stabilized well after the electroweak phase transition such that  $T_s = T_b$. From the condition \eqref{bound_plasma}, the production mechanism works only when
\bea
\frac{f}{\Mpl} \gtrsim 10^{11}  \times \left(\frac{v}{\Lambda_b}\right)^3 \left( \frac{F}{10^{10}\,{\rm GeV} }\right)^{9/2}.
\eea
As it requires a super-Planckian relaxion decay constant, even the Hubble friction can stablize the relaxion after the reheating as discussed in the introduction.

Since we are discussing the hypercharge gauge boson, the condition $\Delta V \leq T_s^4$ to avoid a too much dark radiation can be relaxed. Instead, one might require the condition $\Delta \phi \leq (\Lambda^2 v^2 / \Lambda_b^4) f$ to avoid a too large change of the pre-selected Higgs boson mass. In this case, the initial energy density of relaxion can be as large as $\Lambda^2 v^2$ so that the relaxion energy density dominates the universe before the temperature of the universe is decreased down to $T_s$. In this case, a large portion of the entropy of the current universe originates from the entropy that is relased from the relaxion condensate. Such entropy release from the relaxion condensate could dilute, for instance, bayon number which is generated by high temperature dynamics, or density perturbation which is generated by inflaton fluctuation. Obviously the dynamics of relaxion in such case is more involved, so we leave the detailed study of this case to future work.

\section{further constraints}\label{sec:discussion}

As noticed recently, the relaxion mass $m_\phi\simeq \Lambda_b^2/f$ and the decay constant $f$ can be constrained  by a variety of low energy observables, as well as by astrophysical and cosmological considerations \cite{Choi:2016luu, Flacke:2016szy}. In this work, we assume that the relaxion couples to the Standard Model particles mostly  through the mixing with the Higgs boson, which arises from the barrier potential,
\bea
V_{b} = \mu_b^2 |h|^2 \cos (\phi / f ),
\eea
yielding the relaxion-Higgs mixing angle 
\bea
\theta_{h\phi} \,\sim\, \frac{\Lambda_{b}^4}{m_h^2} \frac{1}{v f}.
\eea
For a relaxion mass $m_\phi \simeq  \Lambda_{b}^2/ f \gtrsim {\cal O}(100\MeV)$, low energy precision measurements such as rare meson decay \cite{Choi:2016luu, Flacke:2016szy} already put severe constraints. Supernova 1987A provides constraints on the lower mass range ${\cal O}(0.1\MeV) \le m_\phi \le {\cal O}(100\MeV)$ \cite{Choi:2016luu}, while $m_{\phi}={\cal O}(1 \keV)$ is constrained by globular clusters \cite{Flacke:2016szy} if the mixing $\theta_{h\phi}$ is as large as ${\cal O}(10^{-9})$. Finally the fifth force experiments can constrain the lighter relaxion with $m_\phi\le {\cal O} (100\meV)$ \cite{Flacke:2016szy}. 
As the heavier relaxion is severely constrained by various observational data, in the following we  focus on the relaxion mass $m_\phi \lesssim 0.1\MeV$ with small mixing angle $\theta_{h\phi} \sim 10^{-9}$, which can be consistent with the existing astrophysical constraints.

\begin{figure*}
\begin{center}
\begin{tabular}{cc}
\hspace{0pt}\includegraphics[scale=0.2]{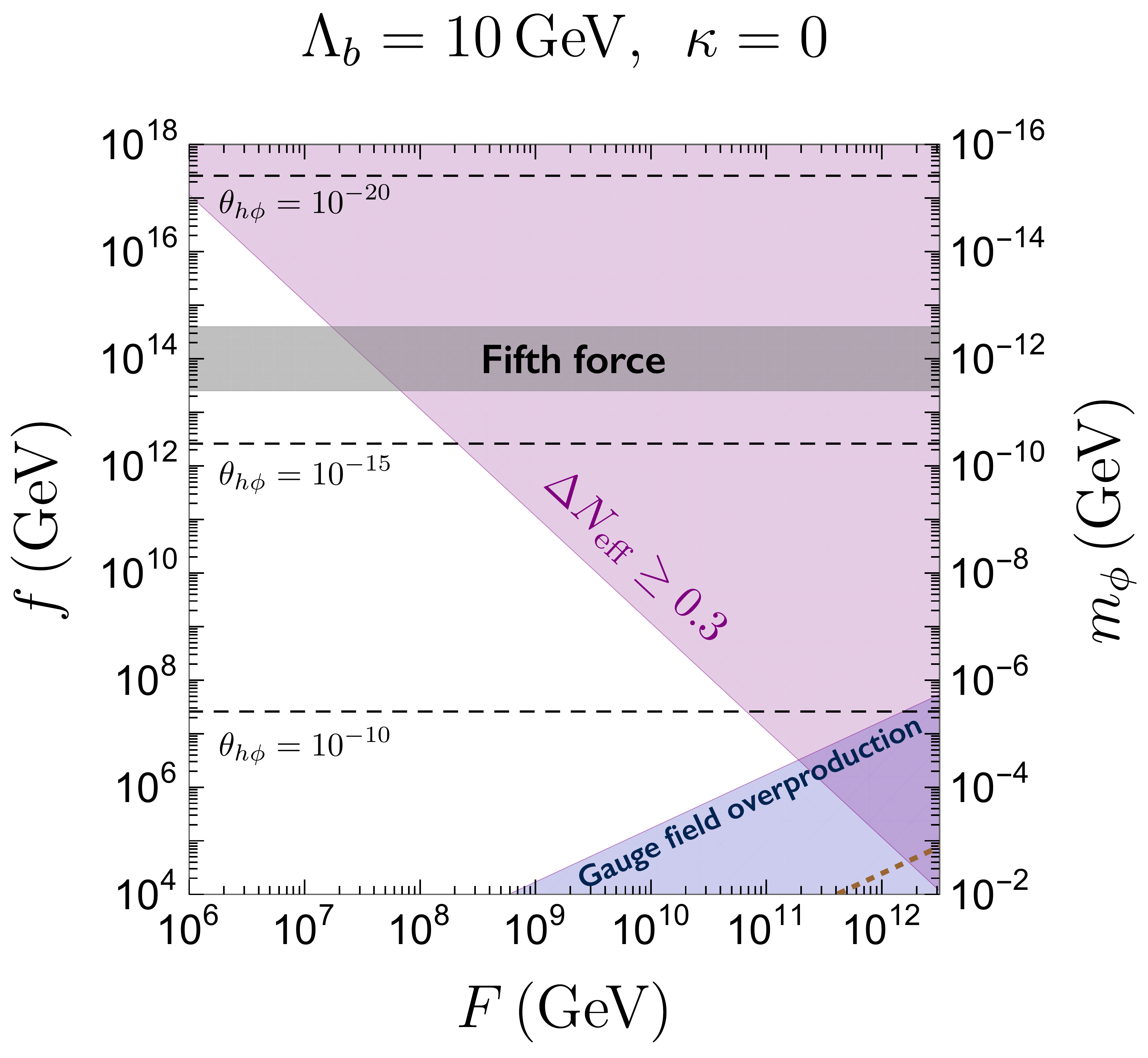} &
\hspace{0pt}\includegraphics[scale=0.2]{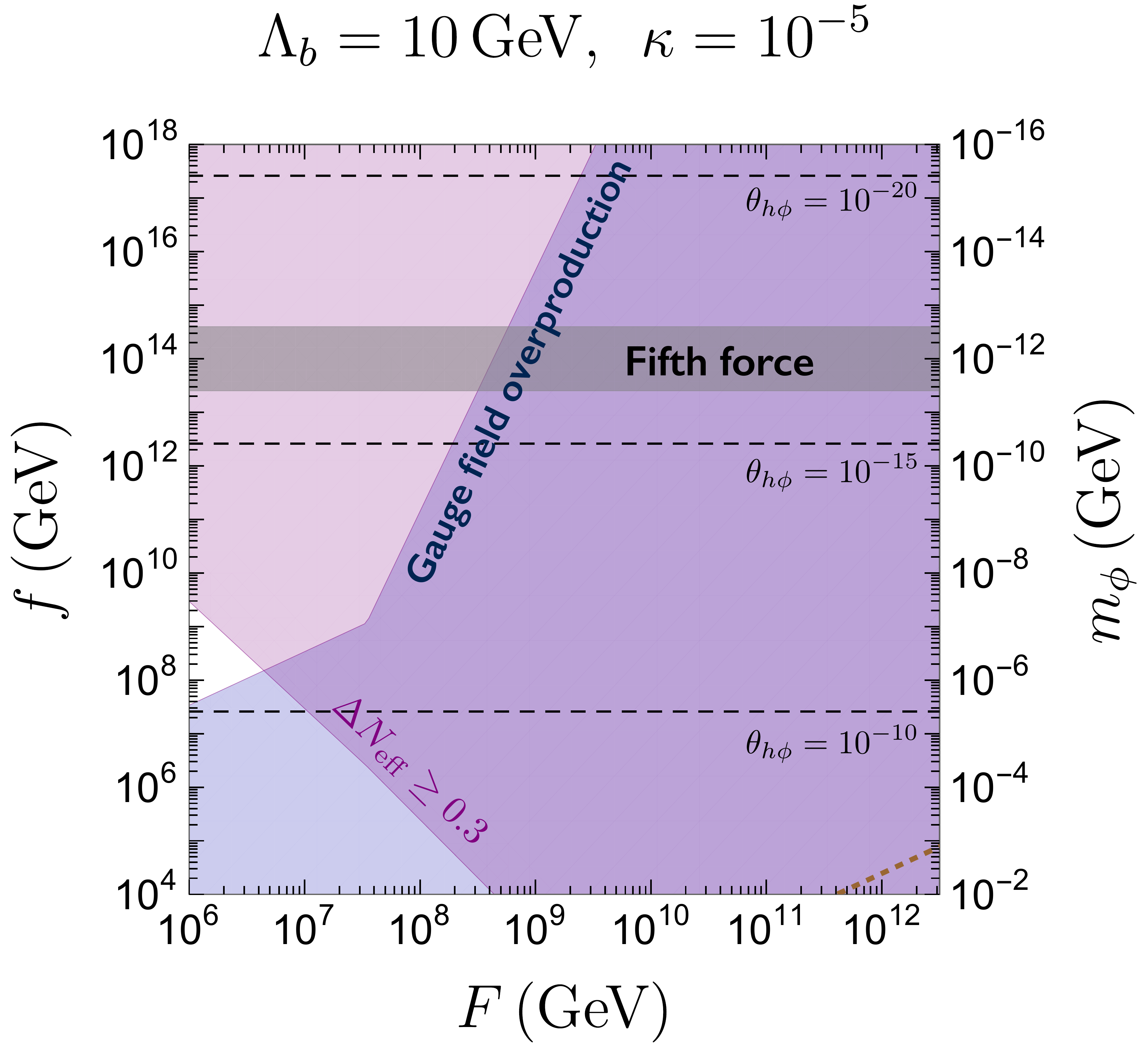} \\
\hspace{0pt}\includegraphics[scale=0.2]{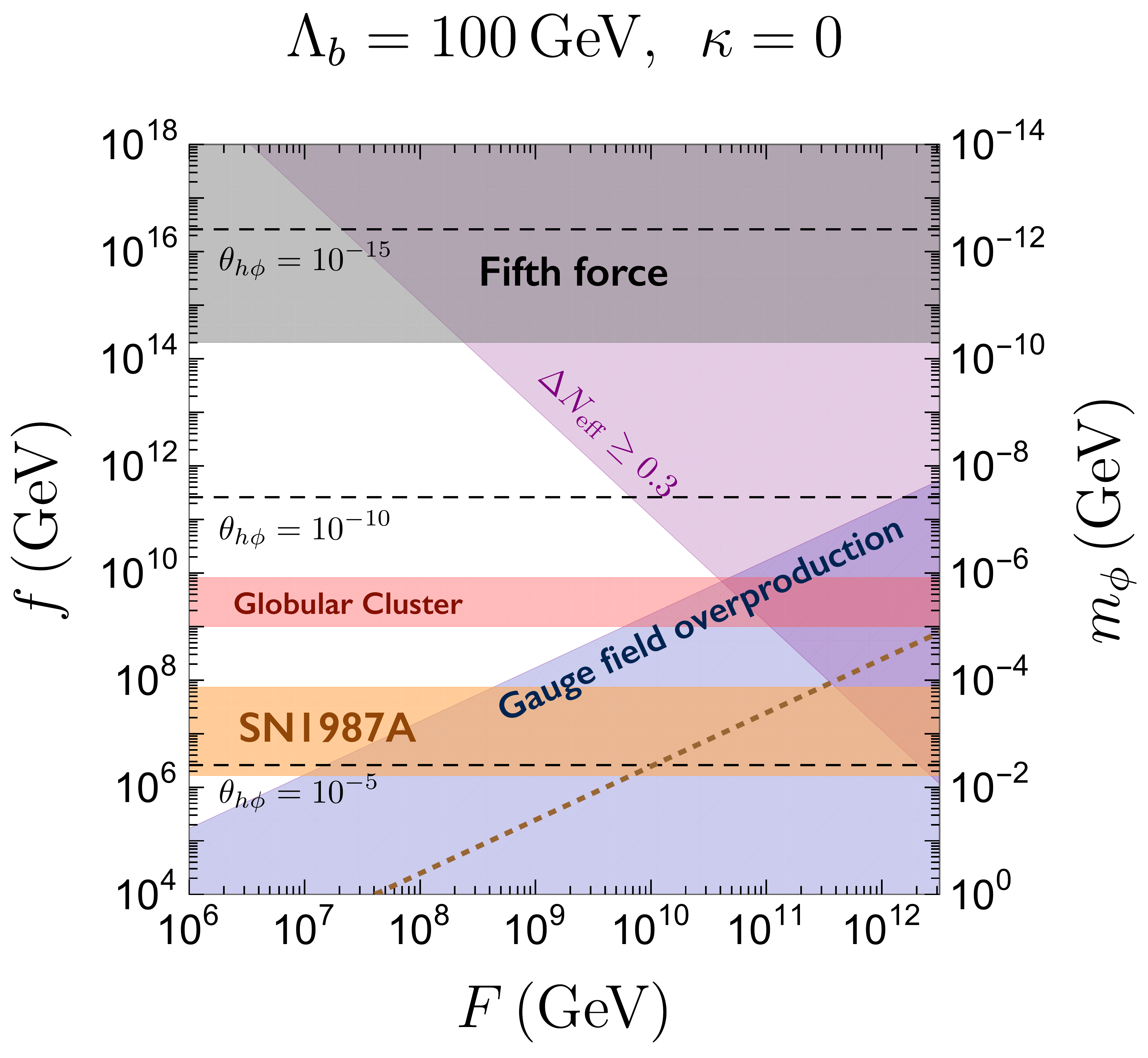} &
\hspace{0pt}\includegraphics[scale=0.2]{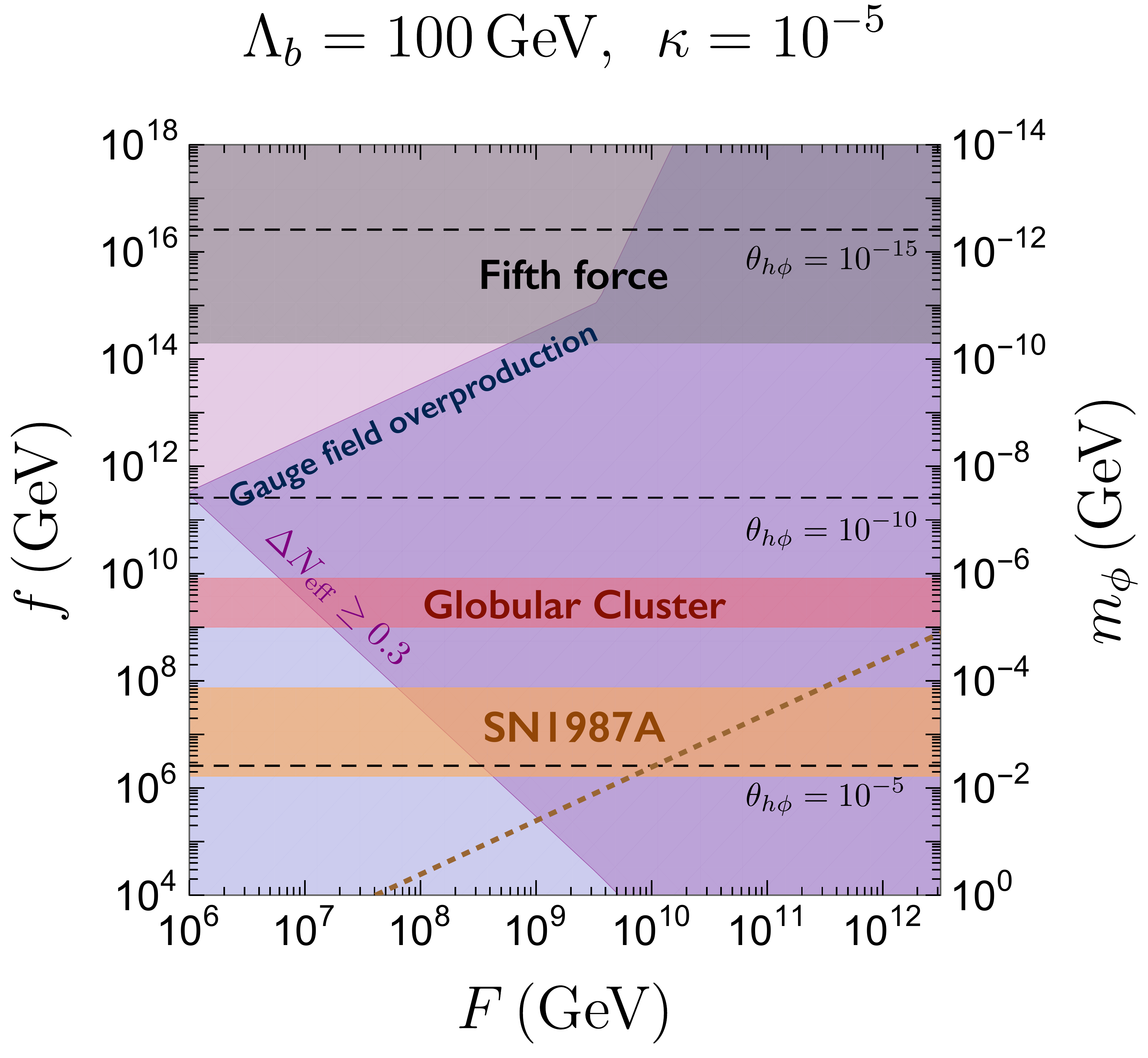} 
\end{tabular}
\end{center}
\caption{Summary of the constraints on model parameters for which the relaxion can be successfully re-stabilized. Uncolored part corresponds to the region satisfying all the available constraints.
Figures in the left column is for the case that
there is no thermal plasma of $U(1)_X$ charged particles, while the right column is for the case with dark plasma having a temperature $T_d= 10^{-5}g^\prime T/g_X$. 
Blue and purple shaded regions are excluded by the overproduction of $U(1)_X$ gauge bosons and the bound on $\Delta N_{\rm eff}$, respectively.  
Above the brown dotted line, relaxions decay dominantly into the $U(1)_X$ gauge bosons, and therefore the bound from late-time relaxion decays in the conventional scenario can be avoided. We also depict other observational constraints in horizontal shaded bands: fifth force search (gray), globular clusters (red), and supernovae (orange). Note that  in the presence of dark plasma the viable parameter region is greatly reduced, so for instance there is only a little viable region for the case with $T_d/T= 10^{-5}g^\prime/g_X$ and $\Lambda_b \geq 10$ GeV.
}
\label{fig:param}
\end{figure*}

In the above, cosmological constraints from the big bang nucleosynthesis, 
cosmic microwave background, 
and extragalactic background light are not included. Although they provide sensitive probes for the mass range ${\cal O}(1\keV) \leq m_\phi \leq {\cal O}(100 \MeV)$ in the conventional relaxion model \cite{Choi:2016luu, Flacke:2016szy}, those cosmological constraints rely on the decay of relic relaxions into the SM particles after the neutrino decoupling. On the other hand, in our scenario with the coupling (\ref{anomalous_coupling}),
relaxions can decay dominantly into the $U(1)_X$ gauge bosons if the coupling $1/F$ is large enough compared for instance to the relaxion-photon coupling $\alpha\theta_{h\phi}/\pi v$ induced by the relaxion-Higgs mixing. Indeed in such case
many of the cosmological constraints discussed in \cite{Choi:2016luu, Flacke:2016szy} can be circumvented as summarized in Fig.~\ref{fig:param}.

Meanwhile, the $U(1)_X$ gauge bosons produced by the late relaxion decays contribute to the dark radiation, which will be discussed in 
 the following. We remind the reader that the present constraint on the number of relativistic degrees of freedom is $N_{\rm eff} = 3.15 \pm 0.23$ \cite{Ade:2015xua}, providing an upper bound on the effective number of neutrino species as
 $\Delta N_{\rm eff}\lesssim 0.3$.

After $t_s$, the relaxion starts to oscillate around the EW vacuum. 
At first, the oscillation is overdamped due to the friction from the gauge field as $t<t_s$. We assume that the duration of the over-damped oscillation is not much larger than the Hubble time. This is automatically satisfied when $f\lesssim \xi F$. On the other hand, if $f\gg \xi F$ without thermal mass, this requires an atypically small initial displacement of the relaxion from the minimum, $\Delta \phi=\mathcal O(\xi F)\ll f$.\footnote{
We note the same subtlety resides in the conventional relaxion scenario. The initial relaxion VEV is not necessarily very close to the EW vacuum since the relaxion mass is smaller than the inflation Hubble rate \cite{Choi:2016luu}. Therefore, at some time after inflation the relaxion starts to oscillate with initial oscillation energy $\simeq\Lambda_b^4$, and can easily dominate the Universe at later time, unless the VEV happens to be atypically close to the minimum.
}
This fine-tune in the initial displacement may be avoided if, for example, the gauge field $X_\mu$ have a tiny (thermal) mass substantially larger than the Hubble rate after $t_s$. In the following, we however assume a tuned initial displacement and restrict ourselves to the setups we adopted in the previous section to make arguments straightforward.

Provided the assumption above,
after settled down at $t_s$, the relaxion field oscillates coherently with an energy
density given by
\bea
\rho_\phi(t) \sim \dot{\phi}^2(t_s) \left( \frac{a_s}{a} \right)^3
\quad \textrm{for} \quad t>t_s. \label{rhophi}
\eea
As we require $\dot{\phi}(t_s) \lesssim \Lambda_{b}^2$ and also $\Lambda_{b}^2 \lesssim {\cal O}(4\pi v^2)$, the energy density of the relaxion condensate is smaller than the radiation energy density  at $t=t_s$. 
Eventually, relaxions decay into the $U(1)_X$ gauge bosons, and the resulting  contribution to $N_{\rm eff}$ depends on the relaxion life time, as well as on the initial relaxion energy density.
As the temperature at the time of relaxion decay is given by $T_{\rm dec} \simeq 1.7 g_*^{-1/4}\sqrt{ \Gamma \Mp}$ with the decay width  $\Gamma = m_\phi^3/ 64 \pi F^2$, 
we find the  contribution from $U(1)_X$ gauge bosons to the relativistic degrees of freedom at $T_{\rm dec}$ is given by
\bea
\Delta N_{\rm eff} = 
\frac{8}{7} \left(\frac{11}{4}\right)^{4/3} 
\frac{g_{*s}(T_{\rm dec})}{g_{*s}(T_s)}
\frac{\rho_\phi (t_s)}{\rho_\gamma(t_s)} \frac{T_s}{T_{\rm dec}}.
\eea

In addition to those in the form of coherent oscillation,  relaxions can be produced thermally from the SM plasma, providing another contribution to $\Delta N_{\rm eff}$. 
For $\theta_{h\phi}$ as small as $10^{-9}$, dominant production channels are the SM particle scatterings producing $\phi$ and gluon, 
which would yield  the relaxion abundance $n_\phi/T^3\sim 3\times 10^{-9}(\theta_{h\phi}/10^{-9})^2$ \cite{Flacke:2016szy}. This results in $\Delta N_{\rm eff}\sim 10^{-6} (\theta_{h\phi}/10^{-9})^{3/2} (F/10^9\GeV) (\Lambda_b/100\GeV)$, which is subdominant compared to the contribution from coherently oscillating relaxion condensate.

So far we have assumed that the $U(1)_X$ gauge bosons are not in thermal equilibrium with the SM plasma. If they are thermalized and remain in thermal equilibrium until the late time, $U(1)_X$ gauge bosons can be as abundant as neutrinos, which would violate the bound on $\Delta N_{\rm eff}$.
On the other hand, in the second example that we have discussed  in the previous section, we take $\kappa \equiv g_XT_d/g'T$ as a free parameter. However, if the dark sector has been in thermal equilibrium with the SM particles after the reheating, the temperature of two sectors should be almost the same except for the small difference coming from entropy boost. As we will see in the following discussion, this is disadvantageous to our scenario. Ignoring the kinetic mixing between the $U(1)_Y$ and $U(1)_X$ gauge bosons, the SM sector and the dark sector interact with each other mostly through
the relaxion; e.g., two SM fermions can annihilate into the $U(1)_X$ gauge bosons mediated by the relaxion. The interaction rate of such a process is given by
\bea
n_f \langle \sigma({ff\rightarrow XX}) \rangle \,\sim\, \theta_{h\phi}^2 \left(\frac{m_f}{v} \right)^2 \frac{T^3}{F^2},
\eea
and the thermalization of the dark sector is possible for the temperature
\bea
T\gtrsim \theta_{h\phi}^{-2} \frac{F^2}{\Mp}  \left(\frac{v}{m_f}\right)^2.
\eea
Note that for $m_\phi \leq {\cal O}(0.1 \, {\rm MeV})$, the relaxion-Higgs mixing angle can be at most ${\cal O}(10^{-9})$. Also  $F\gtrsim {\cal O}(8\pi^2 \Lambda)$ for theoretical consistency. This means that even for a relatively low cutoff scale, the thermalization of  dark sector is  possible only for high reheating temperature, e.g. $T\gtrsim {\cal O}(10^{10} \, {\rm GeV})$ for $\Lambda=10$ TeV.

In Fig.~\ref{fig:param}, we show the viable parameter region for our scenario in the absence (left) and presence (right) of dark plasma of the SM singlet but $U(1)_X$-charged particles. For the purpose of illustration, we adopt the  dark sector temperature $T_d = \kappa T (g'/g_X)$ with $\kappa =10^{-5}$. There are two primary requirements: conditions for successful relaxation, i.e. \eqref{bound} or \eqref{bound_plasma} discussed in the previous section,  and additional constraints for $\Delta N_{\rm eff}\le 0.3$ discussed in this section. Other constraints from astrophysics and terrestrial experiments \cite{Choi:2016luu, Flacke:2016szy} are overlaid in the same plot. First of all, in the presence of dark plasma providing a thermal mass $m_D \gg H$, our scenario works only on a very limited region of parameter space as described in Fig.~\ref{fig:param}. The viable parameter regions shrink even more if we increase $\kappa$. This essentially forbids the $U(1)_X$ gauge boson to be identified as the SM  $U(1)_Y$ gauge boson for the most of parameter space, as stated in the end of previous section. On the other hand, our mechanism works for a reasonably wide range of parameter space in the absence of dark plasma. In particular, $\Lambda_b$ can be as large as ${\cal O}(100\GeV)$ which is preferred in view of the inflationary model building. Such a large $\Lambda_b$ is constrained further by the fifth force experiments and stellar evolution in globular clusters. Our scenario may be tested if the sensitivity of these experiments is significantly improved in the future.

We finally comment on the possible perturbations of the relaxion field around homogeneous background. As in the conventional relaxation scenario \cite{Graham:2015cka}, we assume that the electroweak scale is selected during the long period of inflation. Then the background value of relaxion at the beginning of reheating is nearly homogeneous.
On the other hand, since the production of $U(1)_X$ bosons depends on the gauge field wave number, one may suspect that an inhomogeneity of $\phi$ might be developed consequently. However, it turns out that due to the negative feedback working between the relaxion speed and the gauge field production, the relaxion excursion in homogenous and isotropic background is  stable against  perturbations. Eventually the relaxion field enters into the terminal regime, and thereafter the growth of perturbations is by no means possible as $\dot\phi$ and the resulting gauge field production continuously decay. Except for possible initial perturbations\footnote{
The relaxion may acquire an initial perturbation, which leads to the neutrino density (NDI) isocurvature perturbation with amplitude $\frac{\Delta N_{\rm eff}}{N_{\rm eff}}\frac{\delta\rho_X}{\rho_X}$ in the uniform density slice. The rms of $\frac{\delta\rho_X}{\rho_X}$ is given by the modulation in $N=\log a$ at the onset of friction-less oscillation, and its rms is roughly estimated as $\frac{H_I/2\pi}{\xi F}$, where $H_I$ denotes the inflationary Hubble scale. On the other hand, in order for the relaxion to follow classical evolution during the inflation epoch, $H_I$ is bounded as $H_I< \Lambda_b (\Lambda_b/f)^{1/3}$. Then the rms of NDI perturbations is much smaller than the observational bound~\cite{Ade:2015lrj} in the allowed regions in Fig. \ref{fig:param}.}, 
we thus conclude that sizable perturbations in the relaxion and $U(1)_X$ gauge fields can hardly be produced in our scenario.

\section{Conclusion}\label{sec:conclusion}

In this paper, we examined  if the cosmological relaxation of the Higgs boson mass, which was proposed recently as an alternative solution to the weak scale hierarchy problem, can be compatible with high reheating temperature well above the weak scale. 
As the barrier potential disappears at high temperature, the relaxion  rolls down further after  the reheating, which may ruin the successful selection of the right Higgs boson mass. As it can provide a working scheme over a wide range of model parameters, we focus on the scenario that the relaxion is coupled to a dark $U(1)_X$ gauge boson as $\phi X\tilde X/4F$. In the presence of this coupling, the background relaxion evolution  causes tachyonic instability of the $U(1)_X$ gauge boson, leading to an explosive gauge field production. Then the relaxion is slowed down soon after the gauge field production, and can be re-stabilized by the barrier potential developed at lower temperature around the electroweak scale.

To identify the working parameter region,  we estimate the relaxion excursion after the reheating and impose the condition not to produce too much $U(1)_X$ gauge bosons, as well as other observational constraints on the model parameters.
We have examined this for two different cases. The first is the case that  there is no thermal plasma of $U(1)_X$-charged particles, so no thermal mass of the $U(1)_X$ gauge boson. The second is the case with dark plasma providing a thermal gauge boson mass $m_D> H$. In the first case, the gauge field production is most efficient, and therefore a successful re-stabilization of the relaxion can be achieved over a wide range of model parameters, including
the one with $\Lambda_b={\cal O}(100)$ GeV which is favored in view of the required  inflationary $e$-folding number in (\ref{efolding_1}). 
In the second case, the thermal mass $m_D>H$  suppresses the gauge boson production, and then the relaxion can be successfully re-stabilized  only for a limited parameter range with smaller $\Lambda_b$.

Through this study, we have also shown that $U(1)_X$ can hardly be identified as the $U(1)_Y$ of the SM under the assumption that the universe has been radiation-dominated until when the relaxion is re-stabilized.
If we adopt the possibility that the universe is dominated by the relaxion energy density over certain period, $U(1)_X$ might be identified as the $U(1)_Y$ for a limited range of model parameters. Although an interesting possibility, the analysis for this case is more involved, and we leave it to future work.

\section*{Acknowledgements}
This work was supported by IBS under the project code, IBS-R018-D1. We would like to thank Hye-Sung Lee, Thomas Flacke, Sang Hui Im, Ryusuke Jinno, and Seokhook Yun for useful discussions.

\end{document}